\documentclass[reprint,aps, prl,superscriptaddress]{revtex4-1}

\usepackage{amsmath, amssymb, times, mathrsfs, hyperref, array, bbm}
\usepackage{graphicx}
\usepackage[usenames,dvipsnames]{xcolor}
\usepackage{braket}

\hypersetup{colorlinks=false}

\bibpunct{[}{]}{,}{n}{}{}
\usepackage{ulem}

\newcommand{\en}{\varepsilon}

\renewcommand{\vec}[1]{{\boldsymbol #1}}

\begin{document}
\title{Majorana bound states in topological insulators without a vortex}
\author{Henry F. Legg}
\affiliation{Department of Physics, University of Basel, Klingelbergstrasse 82, CH-4056 Basel, Switzerland}

\author{Daniel Loss}
\affiliation{Department of Physics, University of Basel, Klingelbergstrasse 82, CH-4056 Basel, Switzerland}

\author{Jelena Klinovaja}
\affiliation{Department of Physics, University of Basel, Klingelbergstrasse 82, CH-4056 Basel, Switzerland}

\begin{abstract}
We consider a three-dimensional topological insulator (TI) wire with a non-uniform chemical potential  induced by gating across the cross-section. This inhomogeneity in chemical potential lifts the degeneracy between two one-dimensional surface state subbands.  A magnetic field applied 
along the wire, due to orbital effects, breaks time-reversal symmetry and lifts the Kramers degeneracy at zero-momentum.
If placed in proximity to an $s$-wave superconductor, the system can be brought into a topological phase at relatively weak magnetic fields. Majorana bound states (MBSs), localized at the ends of  the TI wire, emerge and are present for an exceptionally large region of parameter space in realistic systems. Unlike in previous 
proposals, these MBSs occur without the requirement of a vortex in the superconducting 
pairing potential, which represents a significant simplification for experiments.
Our results open a pathway to the realisation of MBSs in present day TI wire devices. 
\end{abstract}

\maketitle

{\it Introduction.} The non-Abelian statistics of Majorana bound states (MBSs)
makes them a promising basis for fault tolerant quantum computation \cite{kitaev2001,Ivanov2001,kitaev2003}. 
Such MBSs were originally 
predicted to appear at the cores of vortices in spinless topological $p_x+i p_y$ superconductors \cite{volovik1988,read2000,alicea2012}. Fu and Kane \cite{Fu2008} proposed that a topological superconducting phase with MBSs localised at the centers of superconducting vortices could also be present at the surface of a three-dimensional topological insulator (TI) in proximity to an $s$-wave superconductor \cite{hasan2010}.

In a related set-up MBSs have been predicted in thin TI wires that are subjected to a magnetic field parallel to the wire and where the phase of the pairing potential of the proximity-induced superconductivity winds around the wire in the form of a vortex \cite{Cook2011}. While still in their infancy, TI wire devices have recently seen significant experimental progress \cite{hong2012,hamdou2013,cho2015,arango2016,jauregui2016,Ziegler2018,Munning2021,Breunig2021}. For instance, the growth of very thin (diameter $\approx 20$ nm) bulk insulating (Bi$_{1-x}$Sb$_x$)$_2$Te$_3$ wires with quantum confined surface states was reported in Ref.~\citenum{Munning2021}. 

Experimentally the requirement of a vortex in the induced pairing potential essentially necessitates a full superconducting shell, a significant fabrication challenge. For the thinnest wires, where the effects coming from the TI bulk states are weakest, this also requires strong magnetic fields $\sim 6$~T. When the superconductor is attached to only one side of the wire, such that a vortex is not expected,  the superconducting gap in the topological phase was shown to be negligibly small \cite{Cook2012,deJuan2014,deJuan2019}, meaning that MBSs have very long localisation lengths and will overlap in realistic finite-length wires. Hence, the prerequisite for a vortex presents a major roadblock to the realisation of MBSs in current TI wire devices. 

In this work,  we propose an alternative protocol for obtaining MBSs in TI wires without the requirement of a superconducting vortex. This is accomplished by a metallic gate placed below the wire and a superconductor attached to the top (as shown in Fig. 1). Such a set-up is very natural since gating will almost always be required to tune the system to the Dirac point \cite{Munning2021}. The non-uniform chemical potential induced by a finite gate voltage breaks inversion symmetry, splitting the doubly degenerate one-dimensional (1D) subbands of the quantum confined TI surface states. A magnetic field applied parallel to the wire opens a gap at momentum $k=0$ due to orbital effects and leaves only a pair of (almost) helical modes at finite Fermi momenta. As a result, when, in addition, the superconductivity in the TI wire is induced by proximity to an $s$-wave bulk superconductor, a topological superconducting phase can be achieved.
In contrast to a TI wire in which inversion symmetry is not broken by gating, this topological phase, induced without the requirement of a vortex, is characterized by a relatively large superconducting gap in the spectrum. As a consequence the topological phase with well localised zero-energy MBSs occupies an exceptionally large  area in  parameter space. In realistic wires the range of chemical potential with MBSs spans several meV at relatively weak magnetic field strengths. Our protocol therefore provides a path to realising MBSs in present day TI wire devices.

\begin{figure}[t]
	\centering
  \includegraphics[width=0.7\columnwidth]{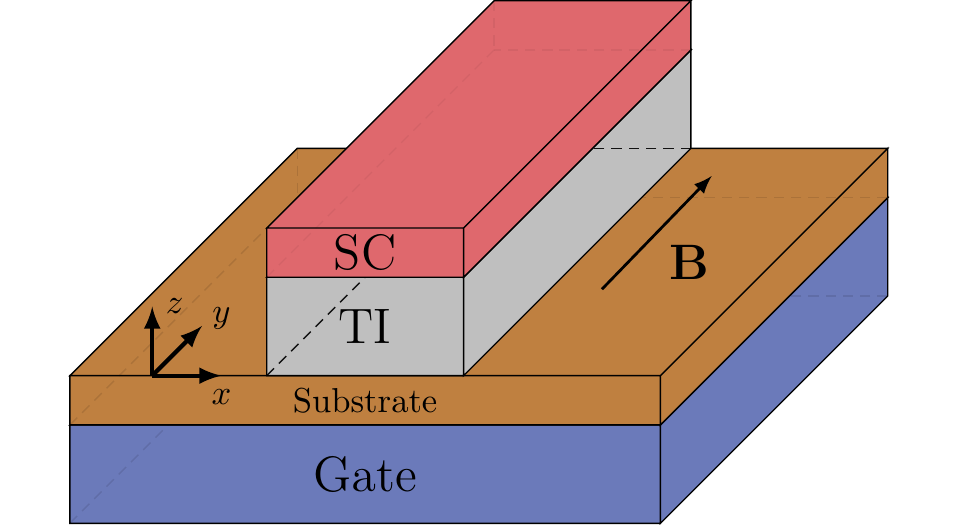}
	\caption{{\bf Gated TI wire:} A metallic gate (blue) induces a non-uniform chemical potential across the TI wire of cross-sectional area $L_x\times L_z$, breaking inversion symmetry and lifting the twofold degeneracy of the surface state subbands with opposite angular momenta. When a pairing potential is proximity-induced by a superconducting layer placed on top of the wire (red) and a magnetic field, $\bf B$ is applied parallel to the wire axis, the system can enter into a topological superconducting phase if the flux produced by $\bf B$ exceeds a critical value. Unlike when no gating is present, within this topological phase MBSs can be achieved without the requirement of a vortex and are well-localised to the ends of the wire due to a large superconducting gap.}
	\label{fig:setup}
\end{figure}

{\it Subband splitting.} We begin with a continuum model for the Dirac surface states of a three-dimensional TI  confined to a cylindrical wire. Due to confinement the surface states form 1D subbands labelled by half-integer angular momentum $\ell=\pm\frac{1}{2},\pm\frac{3}{2},\dots$ around the wire and with an energy dispersion above the Dirac point given by~\cite{si}
\begin{equation}
\epsilon_\ell(k)=\hbar v_F \sqrt{k^2+(\ell-\varphi)^2/R^2},
\end{equation} 
and by $-\epsilon_\ell(k)$ below the Dirac point. Here, $2\pi R$ is the perimeter of the wire cross-section, $k$ is the momentum along the wire, $v_F$ is the Fermi velocity, and $\varphi\equiv\Phi/\Phi_0=B\pi R^2 /\Phi_0$ is the magnetic flux induced by the magnetic field $B$ applied along the wire in units of the fundamental flux quantum $\Phi_0=h/e$. Inversion symmetry requires that the subbands satisfy $\en_{\ell}(k)=\en_\ell(-k)$  and, in the absence of a magnetic field, time reversal symmetry enforces $\en_{\ell}(k)=\en_{-\ell}(-k)$ \cite{deJuan2019,si}. Therefore, when both symmetries are present, subbands are doubly degenerate for angular momenta $\pm \ell$ at all momenta $k$.

We first consider the influence of the gate in the absence of magnetic fields. Since the gate is placed on only one side of the wire, the resulting electrostatic field gives rise to a non-uniform chemical potential 
$\mu(\theta)=\mu_0+\delta \mu(\theta)$ across the wire cross-section \cite{Ziegler2018,Munning2021}, where $\mu_0$ is the average chemical potential and $\theta$ is the angle from the direction normal to the gate (i.e. from the $z$ direction  in Fig.~\ref{fig:setup})\cite{si}. This inhomogeneous chemical potential $\delta \mu(\theta)$ breaks inversion symmetry and therefore lifts the degeneracy of the subbands at finite $k$. Kramers theorem requires that $ k \neq 0$ subbands have opposite spins at opposite momenta and at $k=0$ subbands remain degenerate.

More precisely, within this continuum model, one can obtain the subband splitting by expanding the inhomogeneous component of the chemical potential $\delta\mu(\theta)$ in terms of Fourier cosine harmonics $\delta \mu(\theta)=2\sum_{n=1}^\infty \mu_{n} \cos(n\theta)$. In general, the inhomogeneous potential $\delta \mu(\theta)$ causes finite matrix elements $\mathcal{M}_{\ell \tau}^{\ell' \tau'}(k)$ between the $\ell$ and $\ell'$ subband, where $\tau=\pm$ indicates subbands above (+) or below (-) the Dirac point \cite{Munning2021,si}. When the inhomogeneity is small we can use degenerate perturbation theory for the $\ell>0$ and $-\ell$ subbands and label the resulting subband pair by $\ell$. The $n=2|\ell|$th Fourier component of the inhomogeneous potential $\delta \mu(\theta)$ couples these degenerate subbands and results in a splitting of subbands above the Dirac point
\begin{equation}
\en^{\pm}_{\ell}(k)\approx \epsilon_{\ell} (k) \pm \frac{\mu_{2\ell}k}{\sqrt{k^2+(\ell/R)^2}}.
\label{eq:split}
\end{equation}
As a result, each subband minimum moves from $k=0$ to a new minimum at $\pm k_{so}$, which
can be estimated from Eq.~\eqref{eq:split},
\begin{equation}
k_{\rm so}\approx \mu_{2\ell}/\hbar v_F
\end{equation}
to leading order in the inhomogeneity $\mu_{2\ell}$. The size of the splitting of a given subband pair, $E_{\rm so}(\ell)$, is determined by the change in the subband minimal energy as
\begin{equation}
E_{\rm so}(\ell)=|\epsilon_{\ell}(0)-\en^{\pm}_{\ell}(\pm k_{\rm so})|\approx \frac{\mu_{2\ell}^2}{2\ell \hbar v_F/{ R}}.
\end{equation}
Since inhomogeneities are typically of the order of the subband spacing $\delta\varepsilon \sim \hbar v_F/R \sim 30$ meV \cite{Ziegler2018,Munning2021}, the thinnest  $R \sim 10$ nm experimental TI wires can achieve splitting energies $E_{\rm so}\gtrsim 10$~meV with a corresponding length scale $l_{\rm so}\equiv 1/k_{\rm so}\lesssim 15$~nm. We note that this energy is a very large value when compared to the similar subband splitting caused by Rashba spin-orbit coupling  in semiconducting nanowires, which is typically $E_{\rm so} < 1$~meV in InAs or InSb \cite{prada2020}. 

\begin{figure}[t]
\includegraphics[width=1\columnwidth]{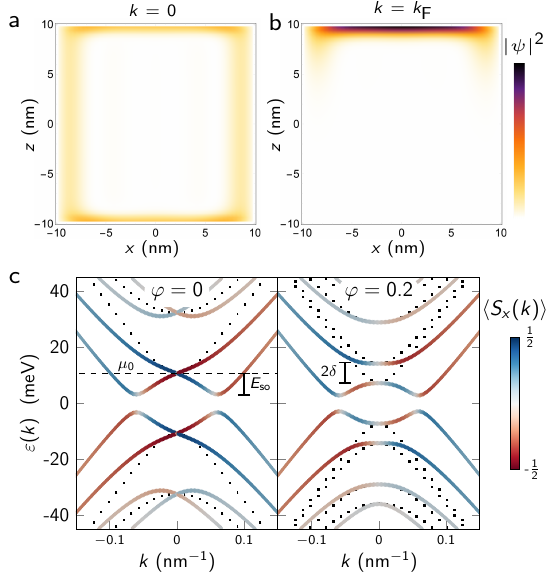}
\caption{
{\bf Wave functions and subband splitting in gated TIs:} The energy spectrum of a $20$ nm $\times$ $20$ nm cross-section of an infinitely long Bi$_2$Se$_3$ TI wire found numerically using the BHZ tight binding model with the chemical potential inhomogeneity given by $\delta \mu(z)=3 \delta\varepsilon (z/L_z) $, where $z$ is measured from the center of the wire, and $\delta\varepsilon\approx 21$ meV \cite{si}. {\bf (a - b)} Probability densities $|\psi_{k}(x,z)|^2$ in the cross-section of the wire. At $k=0$ the state remains uniformly distributed around the wire surface but at $k=k_F$ it is localised close to the top (or bottom) due to the gating. {(\bf c)} The energy spectrum of the gated subbands. Colours indicate the expectation value of spin component  perpendicular to the gate normal:  $\langle S_x (k) \rangle =\int d xdz \bra{\psi_{k}(x,z)}\sigma_x \ket{\psi_{k}(x,z)}$, where $\sigma_i$ are the Pauli matrices encoding the spin in the BHZ model. Left: A large subband splitting occurs due to  gating. Dotted black lines refer to subbands before gating. Right: Applying a magnetic field $B$ along the wire opens a gap of several meV for relatively small fields due to orbital effects. For this thin wire the flux $\varphi=0.2$ corresponds to $B\approx 2$~T and leads to a gap $2\delta(\varphi=0.2)\approx 8$ meV.}\label{fig:subbands}
\end{figure}

The gate induced splitting of subbands is also confirmed numerically (see Fig.~\ref{fig:subbands}) by diagonalising the Bernevig-Hughes-Zhang (BHZ) tight-binding model Hamiltonian in momentum space for a wire with a square cross-section \cite{bernevig2006,zhang2009,Liu2010,schulz2020}. To connect to realistic experimental systems throughout this manuscript we use parameters consistent with the TI Bi$_2$Se$_3$ \cite{Liu2010,si}, choosing wires with a $20$~nm~$\times$ $20$~nm cross-section and the crystallographic axes $c\|\hat{\vec z}$ and $ab$-plane spanning the $\hat{\vec x}$ and $\hat{\vec y}$ directions \cite{Munning2021}. As expected from our considerations above, we find that the subband splitting does not  strongly depend on the exact choice of the inhomogeneity of the chemical potential and so will be generically be present in any TI wire that has been gated \cite{si}. 
Although a perturbative result, as seen in Fig.~\ref{fig:subbands}, the energy spectrum defined by  Eq.~\eqref{eq:split} is still relevant even for inhomogeneities larger than the subband spacing $\delta\epsilon$. In particular, we note that the exact shape of inhomogeneity $\delta\mu(\theta)$ induced by the gate does not influence the form of the splitting in Eq.~\eqref{eq:split} other than through the relative sizes of the harmonics $\mu_{2\ell}$.  Ultimately the magnitudes of these harmonics will depend on the wire cross section, gate geometry, and electrostatic considerations such as the screening of the superconductor, but in general splitting will be largest close to the Dirac point as the largest $\mu_{2\ell}$ will generically occur for small $\ell$. From Eq.~\eqref{eq:split} we also see that there is no change in the energy at $k=0$, which is consistent with the previous observation that such states can be connected by Klein-tunneling through any chemical potential inhomogeneity \cite{Ziegler2018,Munning2021}.

{\it Field induced gap.} We now consider a magnetic field applied parallel to the wire, see Fig.~\ref{fig:setup}. A magnetic field opens a gap at $k=0$, leaving an odd number of pairs of counter-propagating modes at the Fermi-level when the chemical potential is tuned close to a subband crossing (dashed line in Fig.~\ref{fig:subbands}c). Even in the presence of the gate induced inhomogeneity $\mu(\theta)$, due to the Klein-tunnelling effects, the wave function at $k=0$ is uniformly distributed on the wire surface. In contrast,  the wave function at $k_F$ is localised close to the top or bottom of the wire, depending on the sign of the corresponding harmonics $\mu_{2\ell}$ (see Fig.~\ref{fig:subbands}a-b). As a consequence, threading a magnetic flux through the gated TI wire opens a gap at $k=0$ due to orbital effects whereas the states at finite Fermi momenta are largely unaffected by such a flux. The size of the gap is given by $\delta (\varphi) \approx \hbar v_F |\varphi|/R$ \cite{si}. For $R\sim 10$ nm wire this means a gap of $2\delta/B \sim 4$ meV per T can be opened. In fact, $\delta$ increases linearly with wire thickness, meaning thicker wires open a gap significantly faster. For instance, a thicker $R \sim 30$ nm would open a gap of $2\delta/B\sim 12$ meV per T. In semiconductor nanowires, which use the Zeeman effect to open an equivalent  gap at $k=0$, this would require a $g$-factor of over 200. The relevant $g$-factor for Bi$_2$Se$_3$ is $g\sim4$ and so the Zeeman effect in our set-up is negligible compared to orbital effects \cite{Liu2010,si}. Finally, we note that the surface states are spin-polarized in the direction orthogonal to the wire and gate normal (the $x$-axis in Fig.~\ref{fig:setup}) and the spin expectation value in the other two directions is zero, $\langle S_y \rangle=\langle S_z \rangle=0$. Therefore if the chemical potential is tuned inside the gap, this setup could be used as a spin-filter with nearly perfect helical modes \cite{klinovaja2011,braunecker2010,streda2003}, see Fig.~\ref{fig:subbands}c.

From the above considerations, we see that, for our protocol, there is a trade-off between the size of the gap opening at a given field (larger in thicker wires) and the size of the subband splitting (larger in thinner wires). For Bi$_2$Se$_3$ and related TI materials the ideal radius $R$ will likely lie in the range $10-30$~nm. In our numerics, we use a $20$~nm~$\times$ $20$~nm cross-section ($R\approx11.3$~nm) at the lower end of this range since smaller cross-sections are numerically more accessible but this likely exaggerates the magnetic field strengths required for the topological phase. In what follows we will focus on the lowest electron subbands $\ell=\pm1/2$ and measure $\mu$ from the subband crossing at $k=0$. That said, our conclusions remain generally applicable also to higher energy subbands as long as the chemical potential inhomogeneity is strong enough, a fact that may prove useful since the screening of the superconductor will limit the range of average chemical potentials, $\mu_0$, accessible in real experiments. 

\begin{figure}[t]
	\centering
  \includegraphics[width=1\columnwidth]{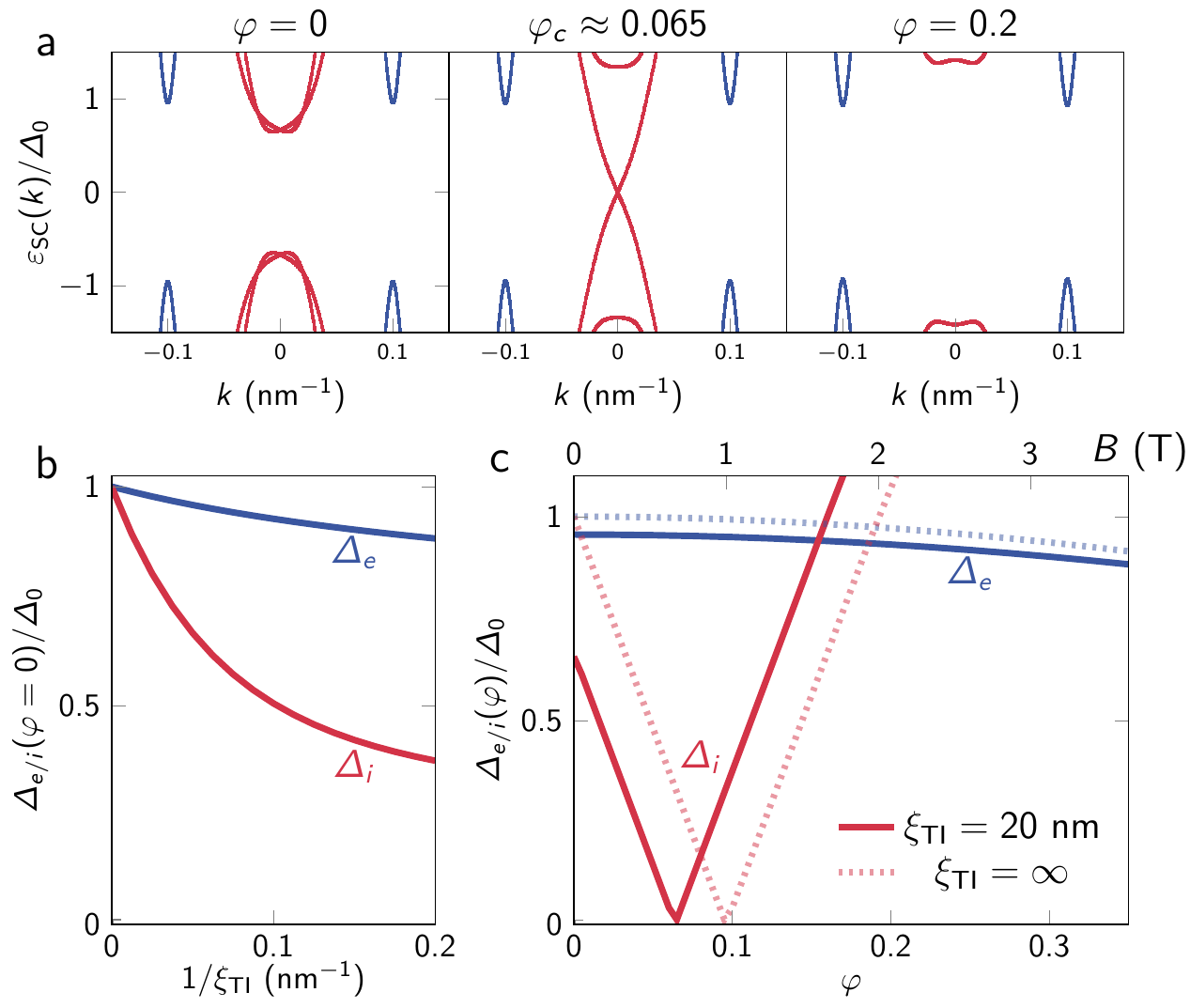}
	\caption{
	{\bf Superconductivity in a gated TI nanowire:} {(\bf a)} Energy spectrum of the gated TI wire from Fig.~2 in the presence of the proximity-induced superconductivity of the strength $\Delta_0\approx1.6$~meV with a coherence length $\xi_{\rm TI}=20$~nm, consistent with the proximity effects in Bi$_2$Se$_3$ thin films due to Nb \cite{flototto2018}. {(\bf b)} The interior superconducting gap $\Delta_i$ is smaller than the exterior gap $\Delta_e$ for a finite coherence length $\xi_{\rm TI}$. {(\bf c)} As the magnetic field is increased the gap at $k=0$ closes and reopens indicating a topological phase transition [for our parameters phase transition occurs at $\varphi_{\rm c}\approx0.065$ ($B\approx650$~mT)]. The exterior gap $\Delta_e \simeq\Delta_0$ stays almost constant. 
	We do not take into account the field dependence of the pairing amplitude and fix $\mu=0$.}
	\label{fig:PBC}
\end{figure}

{\it Topological superconductivity.} Having established the large splitting of the TI wire subbands as well as the presence of the large  magnetic field induced gap $\delta(\varphi)$, the remaining ingredient required for the appearance of MBSs is a proximity-induced superconductivity. We focus on the setup in which an $s$-wave superconductor is placed on the top of the wire and, therefore, the pairing amplitude $\Delta$ is not expected to possess a vortex \cite{deJuan2019}. A simple model for the spatial dependence of the pairing amplitude in such a set-up is
given by $\Delta(z)=\Delta_0 e^{-(L_z/2-z)/\xi_{\rm TI}}$ \cite{flototto2018}, where $\xi_{\rm TI}$ is the superconducting coherence length in the TI. We do not model the reduction of the pairing amplitude by magnetic field but in real systems such an effect will favour thicker wires with a larger gap $\delta(\varphi)$ at a given magnetic field (see above).

\begin{figure}[t]
	\centering
  \includegraphics[width=1\columnwidth]{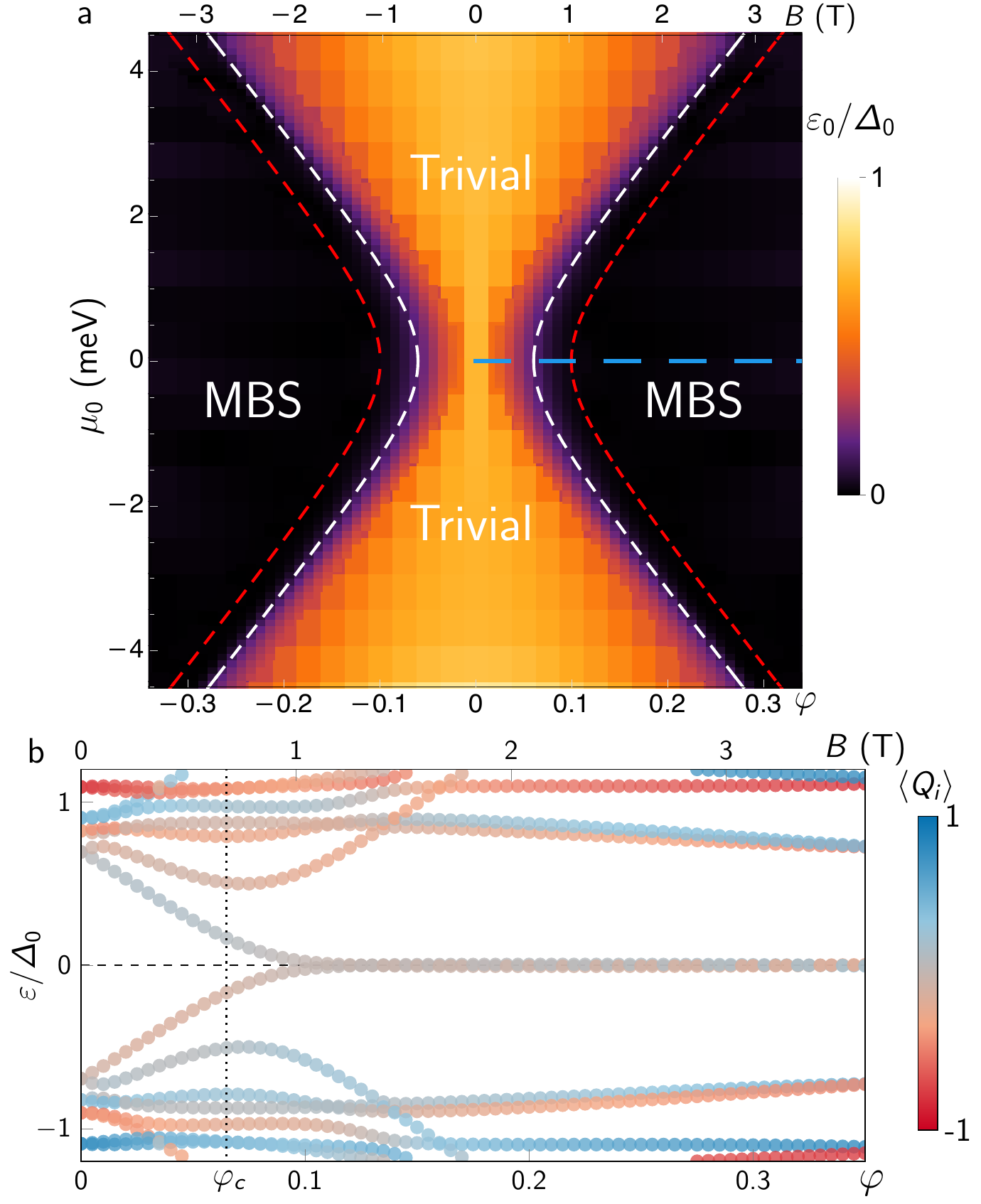}
	\caption{{\bf MBSs in a finite gated TI wire without a vortex:} The same TI wire as in Fig.~3 but of a finite $1000~$nm length. {\bf a)} The lowest energy eigenstates as a function of (average) chemical potential $\mu_0$ and flux $\varphi$. The topological phase transition (white dashed line) corresponds to the closing of the interior bulk gap defined in momentum space. Inside the topological phase (beyond the red dashed line), zero-energy MBSs (black colour) are well separated from the bulk superconducting states, whereas close to the topological phase transition line, their localization length is comparable with the wire length such that they split away from zero energy.
{\bf b)} The lowest part of the energy spectrum [taken along the blue dashed line of (a)], where colours indicate the BCS charge $\langle Q_{i} \rangle=\int d^3r \bra{\psi_{i}(\vec r)}\eta_3\ket{\psi_{i}(\vec r)}$, with $\eta_i$ the Pauli-matrices in particle-hole space \cite{schulz2020}. Zero-energy chargeless MBSs appear in the topological phase when the condition on the localization length is fulfilled [see Eq.~\eqref{eq:fincond}], which for this thin wire happens for a flux $\varphi\approx0.1$ ($B\approx1$~T) which is a bit larger than $\varphi_{\rm c}\approx0.065$ ($B=650$~mT) where the interior bulk gap closes, see Fig.~\ref{fig:PBC}.
}
	\label{fig:OBC}
\end{figure}

Next, we first diagonalise the BHZ tight binding model Hamiltonian in momentum space, where we include the pairing terms of the strength $\Delta$ \cite{si} (see Fig.~\ref{fig:PBC}). To achieve a large superconducting gap, the sign of the chemical potential harmonic $\mu_1$ must be such that the state at $k=\pm k_F$ is localised close to the superconductor. In this case, the exterior superconducting gap, $\Delta_e$, defined around $k=\pm k_F$ (shown in blue in Fig.~\ref{fig:PBC}a),  is essentially unchanged by magnetic field, $\Delta_{e}(\varphi)\simeq \Delta_0$. On the other hand, the interior gap, $\Delta_i$, defined around $k=0$ (shown in red in Fig.~\ref{fig:PBC}a) is smaller $\Delta_i(\varphi=0)<\Delta_0$, even in the absence of  magnetic fields. This can be explained by the fact that the wave functions at finite Fermi momenta are located close to the superconductor, see Fig. 2b.
The amount by which $\Delta_i(\varphi=0)$ is reduced from $\Delta_0$ depends on the coherence length $\xi_{\rm TI}$ (see Fig.~\ref{fig:PBC}b). As the magnetic field is increased, $\Delta_i$ closes  and then reopens, quickly exceeding its zero-field value (see Fig.~\ref{fig:PBC}c).  The closing and reopening of the gap with magnetic field indicates the onset of a topological superconducting phase. The value $\varphi_{\rm c}$ at which the superconducting gap closes is found from the following condition \cite{si}:
\begin{equation}
\delta^2(\varphi_{\rm c})=\mu_0^2+\Delta^2_{i}(\varphi=0)\label{eq:condition}.
\end{equation}
We note that a finite coherence length, which is responsible for smaller values of $\Delta_i$ without diminishing $\Delta_e$, allows one to enter the topological phase at weaker magnetic fields.

{\it Majorana bound states.}  To study properties of MBSs in the topological phase we calculate the energy spectrum of a finite wire \cite{si} and obtain the topological phase diagram, see Fig.~\ref{fig:OBC}.
The white dashed line in Fig.~\ref{fig:OBC}a indicates the topological phase transition line obtained in  Eq.~\eqref{eq:condition}. Inside the topological phase, the MBSs are localized at the ends of the wire and pinned to zero energy, being well-separated from the bulk superconducting states. Close to the phase transition line, the MBSs overlap and split away from zero energy. Thus, generally, the zero-energy MBSs can be observed only if the localization length of MBS  $\xi(\varphi)$, determined by the smallest gap in the spectrum, is smaller than the half of the wire length $L_y$ such that the following criteria is satisfied \cite{rainis2013}:
\begin{equation}
\xi(\varphi)=\frac{\hbar  v_F}{{\rm min}\{\Delta_{i}(\varphi),\Delta_{e}(\varphi)\}}\lesssim L_y/2.\label{eq:fincond}
\end{equation}
This transition line is indicated by the red dashed line in Fig.~\ref{fig:OBC}a. The well-localised zero-energy MBSs persist for a large region of phase space spanning several meV of average chemical potential for fixed finite $B$-fields. As a consequence of this the MBSs are extremely robust against disorder \cite{si}. In stark contrast, without the non-uniformity of chemical potential and with no vortex in the pairing potential, the localisation length condition of Eq.~\eqref{eq:fincond} would not be satisfied for any position in phase space and MBSs would overlap and hybridise, moving away from zero energy.

{\it Conclusions.} We have shown that it is possible to generate Majorana bound states in topological insulator wires without the requirement of a vortex thanks to a large subband splitting of the 1D surface states arising due to a non-uniform chemical potential across the wire cross-section induced by a gate. In our set-up the topological phase occupies a large region of parameter space and MBSs are very robust against disorder. This
 may also allow one to systematically eliminate trivial experimental artefacts, a process which can be difficult in other purported MBS systems. Since a vortex is not required in our protocol it opens a path to MBSs in state-of-the-art TI wire devices.

\begin{acknowledgments}
{\it Acknowledgments.} We acknowledge useful discussions with Y. Ando, A. Rosch, and M. R{\"o}\ss ler.  
This work was supported by the Georg H. Endress Foundation and the Swiss National Science Foundation and NCCR QSIT. This project received funding from the European
Union’s Horizon 2020 research and innovation program (ERC
Starting Grant, Grant No 757725).
\end{acknowledgments}

\bibliography{TI-nws}

\end{document}


\title{Supplemental Material: ``Majorana bound states in topological insulators without a vortex"}
\author{Henry F. Legg}
\affiliation{Department of Physics, University of Basel, Klingelbergstrasse 82, CH-4056 Basel, Switzerland}

\author{Daniel Loss}
\affiliation{Department of Physics, University of Basel, Klingelbergstrasse 82, CH-4056 Basel, Switzerland}

\author{Jelena Klinovaja}
\affiliation{Department of Physics, University of Basel, Klingelbergstrasse 82, CH-4056 Basel, Switzerland}

\maketitle

\renewcommand{\thefigure}{S\arabic{figure}} 

\setcounter{figure}{0}

\section{Subbands and symmetries of TI wire without gating} \label{app:en}
A continuum model for the Dirac surface states confined to a cylindrical wire of radius $R$ has been considered in previous studies of TI wires \cite{Volkov85,Ostrovsky2010,Cook2011,Cook2012,zhang2010,deJuan2019} and we follow a similar derivation in this section. The general Hamiltonian describing TI states confined to the surface  with normal ${\vec n}$ is given by  \cite{Ostrovsky2010}
\begin{equation}
H_{\rm surf}=\frac{\hbar v_F}{2}  {\nabla\cdot {\vec n}}+\frac{v_F}{2}\left[{\vec n}\cdot(\vec p\times \vec \sigma)+(\vec p \times \vec \sigma)\cdot {\vec n}\right], \label{eq:genH}
\end{equation}
where the momentum operator is $\vec p=-i \hbar\nabla$ and $\vec \sigma=(\sigma_x,\sigma_y,\sigma_z)$ is the vector of Pauli matrices, with $\sigma_i$ acting in spin space. 
This Hamiltonian describes a helical system with the spin states being perpendicular to the momentum and normal ${\vec n}$.
For the case of a cylinder along $\hat{\vec y}$ we can write the normal as ${\vec n}=(\sin \theta,0,\cos\theta)$ where $\theta$ is the angle from the $\hat{\vec z}$-direction. The Hamiltonian of the surface states of such a cylindrical TI wire is given by
\begin{align}
H(y,\theta)&=\frac{\hbar v_F}{2R}+i \hbar v_F {\vec  n} \cdot (\vec \sigma \times \nabla)\label{eq:ham1}\\
&=\frac{\hbar v_F}{2R}-i\hbar v_F\left[\sigma_y \frac{\partial_\theta}{R}-(\cos\theta \sigma_x-\sin\theta \sigma_z )\partial_y\right],\nonumber
\end{align}
where we used $\nabla\times {\vec n}=0$ and $\partial_\theta/R=\cos\theta \partial_x -\sin\theta \partial_z$. The solutions are $2\pi$ periodic in the angle $\theta$. The constant term $\frac{\hbar v_F}{2R}$ arises for energies measured from the Dirac point \cite{zhang2010}. Solutions to $H(y,\theta)$ can be found by applying the unitary transformation $U(\theta)=e^{i \theta \sigma_y/2}$ which transforms $H(y,\theta)$ to the simple form
\begin{equation}
\bar{H}(y,\theta)\equiv U(\theta)H(y,\theta)U^\dagger(\theta)=-i \hbar v_F\left(\sigma_y \frac{\partial_{\theta}}{R} - \sigma_x \;\partial_y\right).\label{eq:hamfin}
\end{equation}
Since the transformation satisfies $U(\theta)=-U(\theta+2\pi)$, the boundary conditions for $\theta$ of the eigenstates of the transformed Hamiltonian $\bar{H}(y,\theta)$ are $2\pi$ anti-periodic. We note at this point that the above discussion would be unaffected by the introduction of a non-uniform potential $\delta \mu(\theta)$ since the transformation $U(\theta)$ has no effect on it. Similarly, a Zeeman term $\frac{1}{2} g\mu_B  B\sigma_y$ for a magnetic field $B$ parallel to the wire (see below) would also be unaffected by this transformation.

The solutions of the transformed Hamiltonian of Eq.~\eqref{eq:hamfin} have a simple plane-wave form (up to normalization)
\begin{equation}
\psi_{k\ell\tau}(y, \theta)={\chi_{k\ell\tau}}e^{i(k y +\ell \theta)},\label{eq:wvf}
\end{equation}
where $k$ is the momentum along the wire, ${\chi}_{k\ell\tau}$ is the spinor encoding the chirality, $\tau=\pm$ indicates whether the state is above or below the Dirac point, and $\ell=\pm\frac{1}{2},\frac{3}{2}\dots$ is the angular momentum around the wire which is a half integer due to the anti-periodic boundary conditions in $\theta$. As in Eq.~(1) of the main text, above the Dirac point ($\tau=+$), these states have energies
\begin{equation}
\en_{\ell}(k)=\hbar  v_F \sqrt{k^2+(\ell/R)^2},
\end{equation}
and below the Dirac point ($\tau=-$) have energies $- \en_{\ell}(k)$. The spacing between subbands is given by $ \delta \varepsilon=\hbar v_F /R$.
The normalized eigenspinors are given by
 \begin{equation}
{\chi}_{k\ell\tau}=\frac{1}{\sqrt{2}}\left(\tau,\frac{i \ell/R - k}{\sqrt{k^2+(\ell/R)^2}}\right).\label{eq:eigenspin}
\end{equation}

\section{Matrix elements due to non-uniform chemical potential} \label{app:matelem}
The Hamiltonian $\bar{H}(y,\theta)$ describing surface states of the TI wire [see Eq.~\eqref{eq:hamfin}] has an inversion symmetry of the full three-dimensional space, i.e.  $\bar{H}(y,\theta) =\sigma_y \bar{H}(-y,\theta+\pi) \sigma_y$, which enforces that the energies satisfy $\en_\ell(k)=\en_\ell(-k)$. This inversion symmetry is broken by the non-uniform chemical potential $\delta\mu(\theta)$ induced by the gate. The system also has a time-reversal symmetry, $\bar{H}(y,\theta)=\sigma_y \bar{H}^*(y,\theta) \sigma_y$, which requires $\en_\ell(k)=\en_{-\ell}(-k)$. The combination of these symmetries ensures degeneracy of the subbands for all momenta $k$ \cite{deJuan2019}.

When inversion symmetry is broken by a non-uniform chemical potential $\delta \mu(\theta)$,   matrix elements $\mathcal{M}_{\ell \tau}^{\ell' \tau'}(k)$ 
are induced
which connect states with  different angular momenta $\ell$ and $\ell'$ for potentially different $\tau$ and $\tau'$. These matrix elements result in a subband-splitting, as discussed in the main text. To obtain this splitting analytically, we start with the general form of the matrix elements between the eigenstates $\psi_{k\ell\tau}$ and $\psi_{k'\ell'\tau'}$ that are given by
\begin{align}
\mathcal{M}_{\ell \tau}^{\ell' \tau'}(k)= \bra{\psi_{k\ell\tau}}\delta\mu\ket{\psi_{k\ell'\tau'}}&=  {\chi^\dagger_{k\ell\tau}} {\chi_{k\ell'\tau'}} \int_0^{2\pi} \frac{d\theta}{2\pi} e^{i(\ell'-\ell)\theta}\delta \mu(\theta) \nonumber \\
&= \frac{\mu_{|\ell-\ell'|}}{2}   \left(\tau \tau' + \frac{ (k +i \ell/R )( k - i \ell'/R )}{\sqrt{k^2+(\ell/R)^2}\sqrt{k^2+(\ell'/R)^2}} \right),\,\,\,\, \ell\neq \ell' \, \label{eq:matelem}.
\end{align}
As in the main text, $\mu_{n}$ is the $n$th  Fourier cosine component, $\delta\mu(\theta)=2\sum_{n=1}^\infty \mu_n \cos(n\theta)$, where the geometry of our setup means that the non-uniform potential is equal for the two points $(x,z)$ and $(-x,z)$ on the TI surface or equivalently $\delta\mu(\theta)=\delta\mu(-\theta)$.

Taking $\ell>0$, in the special case of the degenerate subbands $\ell'=-\ell$ with $\tau=\tau'$, this matrix element reduces to
\begin{equation}
\mathcal{M}_{\ell \tau}^{-\ell \tau}(k)=\frac{ \mu_{2\ell} k}{k-i \ell/R}.\label{eq:matelem}
\end{equation}
Assuming that the maximum size of all $\mu_n$ is much smaller than the subband spacing $ \delta\varepsilon$, we can use degenerate perturbation theory to find the energy spectrum. For the bands above the Dirac point, this gives  
\begin{equation}
\en^{\pm}_{\ell}(k) \approx \en_{\ell} (k) \pm \frac{\mu_{2\ell}k}{\sqrt{k^2+(\ell/R)^2}}\label{seq:bands},
\end{equation}
and similarly $-\en^{\pm}_{\ell}(k)$ below the Dirac point [as in Eq.~(3) of the main text]. We find that such splitting between the two degenerate subbands is present for a general form of chemical potential inhomogeneities $\delta\mu(\theta)$ (see Fig.~\ref{sfig:splittingstr}) and not strongly dependent on the exact form of inhomogeneity (see Fig.~\ref{sfig:Splittings} and discussion below). The  momentum $\pm k_{\rm so}$ of the band minimum can be found by solving $\partial_k \en^{\pm}_{\ell}(k)|_{k_{\rm so}}=0 $. One also obtains  the size of the subband splitting defined as $E_{\rm so}(\ell)=|\epsilon_{\ell}(0)-\en^{\pm}_{\ell}(\pm k_{\rm so})|$. Expansion of $k_{\rm so}$ and $E_{\rm so}$ to leading order in $\mu_{2\ell}$ gives Eqs.~(4) and (5) of the main text, respectively. 

For a given TI wire device, to obtain the potential $\delta\mu(\theta)$, one needs to consider the electrostatics of the device \cite{Ziegler2018}. In general $\delta\mu(\theta)$ is a result of both a non-uniformity in charge density, related to the geometric capacitance of the wire, and the direct electrostatic potential, related to the quantum capacitance of the wire. Far from the Dirac point the charge density is most important but quantum capacitance effects can become relevant in the low density region close to the the Dirac point. 

\section{Magnetic field parallel to the wire}
A magnetic field applied parallel to the wire (along the $\hat{\vec y}$ axis in our coordinate system), generally, leads to both orbital and Zeeman contributions to the Hamiltonian. An additional Zeeman term $H_{Z}=\frac{1}{2}g_\| \mu_B B \sigma_y$ added to the Hamiltonian  Eq.~\eqref{eq:ham1} will be unaffected by the unitary transformation $U(\theta)$ we used to derive $\bar{H}(y,\theta)$. This is also true for the orbital component which can be included by minimal coupling, $\vec p \rightarrow \vec p- e \vec A$, with the vector potential $\vec A$ in Eq.~\eqref{eq:ham1}. Using the symmetric gauge $\vec A=\frac{B}{2}(z,0,-x)=\frac{\hbar \varphi}{e R}(\cos\theta,0,-\sin\theta)$ -- which is also unaffected by $U(\theta)$ -- allows us to write
\begin{equation}
\bar{H}(y,\theta)+H_Z=\hbar v_F\left(\sigma_y \left(\frac{-i \partial_{\theta}-\varphi}{R}+\frac{\frac{1}{2}g_\| \mu_B B}{\hbar v_F}\right) +i \sigma_x \partial_y\right),\label{eq:bdep}
\end{equation}
where $\Phi=\pi R^2 B$ is the magnetic flux through the wire, with $\varphi=\Phi/\Phi_0$ the dimensionless flux phase and $\Phi_0=h/e$ the flux quantum, as in the main text. Therefore, a parallel magnetic field can be included by replacing of the angular momentum $\ell \rightarrow  \lambda_\ell \equiv \ell-\varphi +\frac{\frac{1}{2} g_{\|} \mu_B B}{\hbar v_F/R}$. We note that, in this effective low-energy theory, the orbital and spin contributions of a parallel magnetic field add up into one term  with the orbital term usually being dominant \cite{klinovaja2015}.

As discussed in the main text, the field lifts the degeneracy of the modes at $k=0$. To find the size of the gap, we can solve the Hamiltonian for a TI wire surface in the presence of a magnetic field, Eq.~\eqref{eq:bdep}, using  the wave functions $\psi_{k\lambda_\ell\tau}(y,\theta)$ from Eq.~\eqref{eq:wvf}, where we have replaced $\ell \rightarrow  \lambda_\ell$. The energy spectrum is given by $\pm \varepsilon_{\lambda_\ell}(k)$ above (+) and below (-) the Dirac point. At $k=0$, the gap is given by
\begin{equation}
\delta(\varphi)=|\varepsilon_{\lambda_\ell}(0)-\varepsilon_{\lambda_{-\ell}}(0)|/2=\hbar v_F |\lambda_\ell-\lambda_{-\ell}|/2R=|2\hbar v_F \varphi/R - g_\|\mu_B B|/2\approx \hbar v_F |\varphi |/R,
\end{equation}
where the approximation uses the fact the Zeeman contribution is typically much smaller than the orbital flux contribution. For instance, in the $ab$-plane of Bi$_2$Se$_3$ the $g$-factor is $g_\|\sim 4$ and hence the Zeeman contribution $ g_{\|} \mu_B\sim 0.2$ meV/T can be safely neglected in comparison to the orbital contribution $2\hbar v_F |\varphi|/R \gtrsim 4$ meV/T. In tight-binding models (see below) and real systems the finite penetration of the wave function into the TI cross-section requires that $\varphi$ be replaced by the effective flux $\bar{\varphi}$ that, for example, governs the Aharonov-Bohm period and is slightly smaller than $\varphi$. For our parameters and our $20$~nm~$\times$ $20$~nm cross-section $\bar{\varphi}\approx 5 \varphi/6$.

Next, we show that the size of the splitting between two initially degenerate subbands at $k=0$ induced by the magnetic field stays the same even  if the non-homogeneity of the chemical potential is included.  The matrix elements between {\it different} subbands  $\mathcal{M}_{\lambda_\ell \tau}^{\lambda_{\ell'} \tau'}(k=0)$ are given by 
\begin{align}
\mathcal{M}_{\lambda_\ell \tau}^{\lambda_{\ell'} \tau'}(k=0)=  
\frac{\mu_{|\ell-\ell'|}}{2}  [\tau \tau' + {\rm sgn} ({\lambda_\ell  \lambda_{\ell'}}) ].
\end{align}
We focus on the case of weak  magnetic fields $\varphi<1/2$ such that $\lambda_{\ell}$ and $\lambda_{\ell'} $ do not change  sign as a function of the magnetic field. Thus, this matrix element stays equal zero between two subbands ($\ell, \tau)$ and $(-\ell, \tau)$) of our main interest. This means that the gap induced by the magnetic field is not altered by gating and is given by $\delta=\hbar v_F |\varphi|/R$. We confirm this numerically. For instance, in Fig.~2c of the main text, we show that $\delta(\varphi=0.2)$ is essentially identical for the system with and without a non-uniform chemical potential.

\section{Topological superconductivity transition}

In this section, we derive the criterion for the critical flux $\varphi_{\rm c}$ that defines the phase transition to topological superconductivity. The topological superconductivity arises due to the competition between two gap opening mechanisms at $k=0$: namely, between the superconducting gap $\Delta_{i,\ell}$ and the magnetic field gap $\delta_\ell (\varphi)$ for a general subband $\ell$. In what follows, we assume that the uniform part of the chemical potential $\mu_{0,l}$ is calculated from the degeneracy point at $k=0$ for two subbands $\pm \ell$ and treat both $\Delta_{i,\ell}$ and $\delta_\ell(\varphi)$ perturbatively, $\delta_\ell(\varphi),\Delta_{i,\ell} \ll E_{so,\ell}$. In this case, linearizing the spectrum around $k=0$, see Fig.~2c in the main text, the effective low-energy Hamiltonian for the two interior modes, $R_{i,\ell}(y)$ (right-moving field) and $L_{i,\ell}(y)$ (left-moving field), can be written as 
\begin{equation}
H_{\ell}=\hbar {\tilde v}_{F,\ell}k\rho_z \eta_z  +\Delta_{i,\ell} \eta_x +\delta_\ell(\varphi)\rho_x\eta_z -\mu_{0,\ell}\eta_z\, ,
\label{eq:SCHam}
\end{equation}
where $\eta_{x,y,z}$ are the Pauli matrices in particle-hole space, $\rho_{x,y,z} $ are the Pauli matrices acting in the right- and left-moving field space.
The Fermi velocity ${\tilde v}_{F,\ell}$ around $k=0$,  for the cylindrical TI wire, is given by $\hbar {\tilde v}_{F,\ell} = |\mu_{2l}| R/ \ell $.
From Eq. (\ref{eq:SCHam}) it follows that the interior gap closes and reopens at $k=0$ for  $\delta_\ell^2(\varphi_{\rm c})=\mu_{0,\ell}^2+\Delta_{i,\ell} ^2 $. This gives the criterion for the topological phase transition, as in Eq.~(6) of the main text for $\ell=1/2$. 

\section{BHZ tight-binding model without superconductivity} \label{app:tb}
We perform numerical simulations using a BHZ tight binding model without and with a superconducting pairing term \cite{deJuan2019,bernevig2006,liu2010,schulz2020}. Electrostatic factors mean that ultimately the cross section of the wire and gate geometry will impact the relative sizes of the Fourier harmonics, for instance a hexagonal cross-section would result in a large $\mu_6$ Fourier component \cite{Munning2021}. That said, from our analytic treatment we do not expect the cross-section to impact the form of the splitting for a given subband other than through the differing $\mu_n$ and so we choose a square cross-section for our numerical calculations throughout. That we find subband-splittings for square cross-section wire that are consistent with our analytic calculations for a perfect cylindrical wire further validates the fact that the exact cross-section of a given wire is essentially irrelevant as long as the harmonics $\mu_n$ of the non-uniform chemical potential are the same.

\begin{figure}[t]
\centering
\includegraphics[width=0.7\columnwidth]{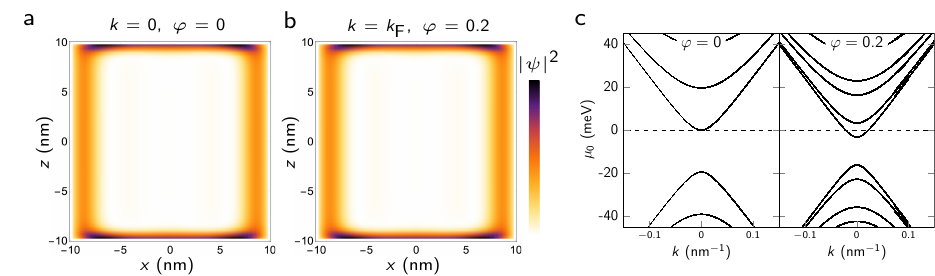}
\caption{{\bf Wave functions and energy spectrum for a TI wire with uniform chemical potential:} The same wire as in Fig.~2 of the main text but without the non-uniformity in the chemical potential. Here, in addition, the bottom of the lowest subband is tuned to zero energy. {(\bf a - b)}  In stark contrast to a gated wire (see Fig.~2 of the main text) the wave functions are not localised on a specific side of the TI wire at finite $k_F$. {(\bf c)} The subbands of the TI wire with a uniform chemical potential are doubly degenerate for $\varphi=0$ and have an energy spacing $\delta \varepsilon \approx 21$~meV.}\label{sfig:uniform}
\end{figure}

We start with the BHZ lattice model of the TI wire with the cross-section $L_x \times L_z$  in the absence of superconductivity. In momentum space, the Hamiltonian  is defined as \cite{deJuan2019,bernevig2006,liu2010,schulz2020} 
\begin{align}
H(k)=\sum_{n=1,m=1}^{L_x/a_1,L_z/a_3} {\bf c}^\dagger_{n,m,k}\cdot \{M(k)\tau_x+\frac{A_2}{a_2} \sin(k a_2) \tau_x \sigma_x  -\mu_{n,m} \}\,{\bf c}_{n,m,k} \nonumber \\
+\sum_{n=1,m=1}^{L_x/a_1-1,L_z/a_3} \left\{{\bf c}^\dagger_{n+1,m,k}\cdot \left\{\frac{B_1}{a_1^2}\tau_x + \frac{i A_1}{2 a_1} \tau_x   \sigma_z \right\}e^{i \phi^x_{m} }  {\bf c}_{n,m,k} +\text{H.c.} \right\} \nonumber \\
+\sum_{n=1,m=1}^{L_x/a_1,L_z/a_3-1} \left\{{\bf c}^\dagger_{n,m+1,k}\cdot \left\{\frac{B_3}{a_3^2}\tau_x + \frac{i A_3}{2 a_3} \tau_y \right\}e^{i \phi^z_{n} } {\bf c}_{n,m,k}+\text{H.c.}\right\},\label{seq:BHZ} 
\end{align}
where $M(k)=M_0-2\frac{B_2}{a_2^2}\cos(k a_2)+2 \left(\frac{B_1}{a_1^2}+\frac{B_2}{a_2^2}+\frac{B_3}{a_3^2}\right)$,  with $A_i$ and $B_i$, $i=1,2,3$, being the BHZ parameters. Here,   ${\bf c}^\dagger_{n,m,k}=(c^\dagger_{+,\uparrow},c^\dagger_{-,\uparrow},c^\dagger_{+,\downarrow},c^\dagger_{-,\downarrow})_{n,m,k}$ is a 4-vector with $c^\dagger_{\pm,\uparrow/\downarrow}$ describing the creation of an electron + (hole -) with spin $\uparrow/\downarrow$ on site $n=x/a_1,m=z/a_3$ and with momentum $k$ along the wire axis in $y$ direction. The Pauli matrices $\tau_{x,y,z}$ act in electron-hole space and $\sigma_{x,y,z}$ in spin space. 
The Peierls phases $\phi_m^x$ and $\phi_n^z$ encode the orbital component of the magnetic field $B$, as above (and also in Ref.~\cite{deJuan2019}) we choose the symmetric gauge $\vec A=\frac{B}{2}(z,0,-x)$ with the origin corresponding to the center of the wire. Explicitly, this means we choose the Peierls phases as 
\begin{equation}
\phi^x_m=\frac{a_1\pi \varphi }{L_x L_z} \left(a_3 \left(m-\frac{1}{2}\right) -\frac{L_z}{2 }\right), \qquad\phi^z_n=-\frac{a_3\pi \varphi }{L_x L_z} \left(a_1 \left(n-\frac{1}{2}\right) -\frac{L_x}{2 }\right)\, ,
\end{equation}
where now $\varphi =BL_x L_z/\Phi_0$.

\begin{figure}[t]
\centering
\includegraphics[width=0.7\columnwidth]{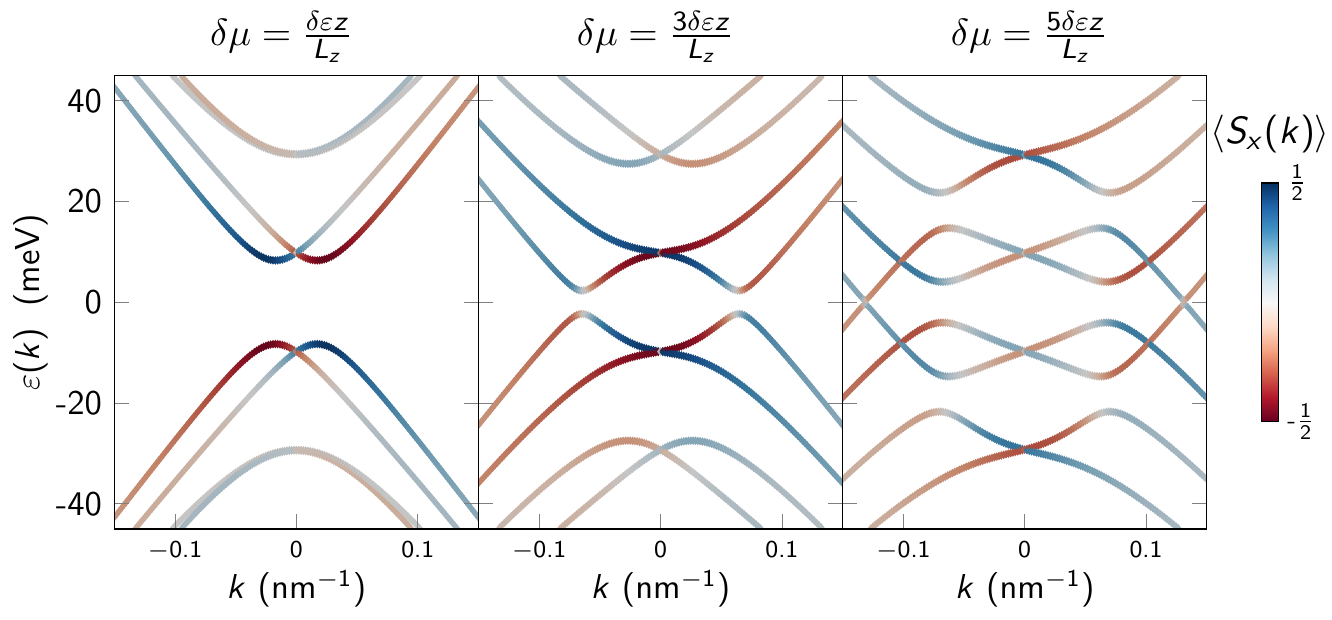}
\caption{{\bf Band splittings for different strengths of non-uniform potential:} The energy spectrum of a Bi$_2$Se$_3$ wire with a $20$ nm$\times20$~nm cross-section for three different strengths of  non-uniformity of chemical potential $\delta\mu$, calculated using the BHZ tight binding model for a wire in momentum space [see Eq.~\eqref{seq:BHZ}]. We find that the perturbative result for the band-splitting given in Eq.~\eqref{seq:bands} is valid for large values of the inhomogeneity $\delta \mu (\theta)$, even when it is larger than the subband spacing $ \delta\varepsilon\approx 21$~meV, and only deviates from the perturbative result when the splitting energy of a given subband is too large $E_{\rm so}\gtrsim \delta\varepsilon/2$.}\label{sfig:splittingstr}
\end{figure}

Throughout the main text and this supplemental material we use the parameters for Bi$_{2}$Se$_{3}$ from Ref.~\cite{liu2010} [Table IV] these are:  $A_1=3.33$ eV\AA, $A_2=\alpha 3.33$ eV\AA, $A_3=2.26$  eV\AA, $M_0=-0.28$ eV,  $B_1=B_2=44.5$ eV\AA$^2$, and $B_3=6.86$ eV\AA$^2$. We always use $\alpha=1$  for wires in momentum space but adjust $\alpha$ to enable simulation of long finite systems (see below) and we choose the crystallographic $c$-axis of Bi$_{2}$Se$_{3}$ parallel to $\hat{\bf z}$. We also always choose lattice constants $a_1=a_2=50/3$ nm in the $\hat{\vec x}$ and $\hat{\vec y}$ directions and $a_3=25/3$ nm in the $\hat{\vec z}$ direction. We find that these are the smallest values such that lattice spacing effects are negligible, this makes our $20$ nm$\times 20$ nm cross-section to be modelled as a system consisting of $12\times24$ lattice sites in the $xz$-plane. Using these parameters for our $20$ nm$\times 20$ nm cross-section  we find $\delta\varepsilon = 2\pi\hbar/(2 L_x/v_x + 2L_z /v_z)\approx 21$~meV, which is a generalised version of $\delta\varepsilon=\hbar v_F/R$ for a square cross-section with different velocities in the $x$ and $z$ directions. These velocities are obtained from the BHZ model parameters using $v_{x}=A_{1}/\hbar$ and $v_{z}=A_{3}/\hbar$. As shown in Fig.~\ref{sfig:uniform}, when the chemical potential is uniform,  the wave function is also uniformly around the wire surface for both Fermi momenta at $k=0$ and at $k=\pm k_F$.  This is very different when the gated TI wire, in which, for finite Fermi momenta, the wave function is more localised on one side of the wire (see Fig.~2 main text).

\begin{figure}[b]
\centering
\includegraphics[width=0.7\columnwidth]{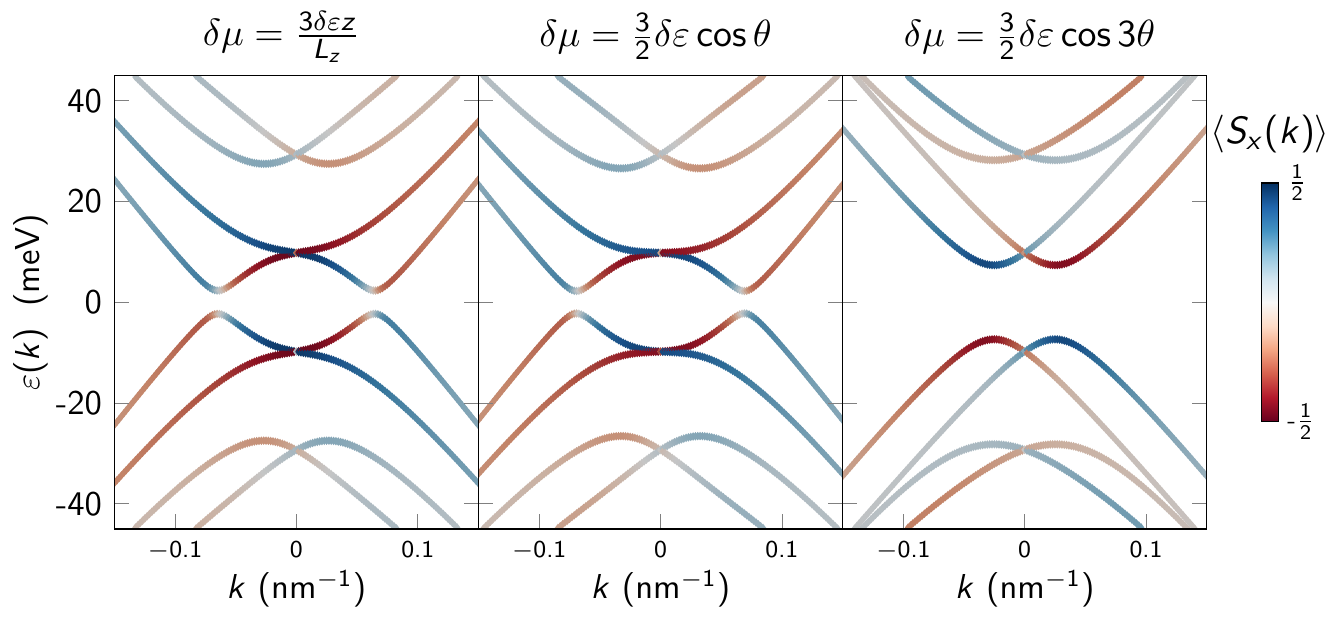}
\caption{{\bf Band splittings for different shapes of non-uniform potential:} The energy spectrum of a Bi$_2$Se$_3$ wire with a $20$ nm$\times20$~nm cross-section for three different non-uniformities of chemical potential, calculated using the BHZ tight binding model for a wire in momentum space [see Eq.~\eqref{seq:BHZ}]. The exact shapes of non-uniform potentials used are listed at the top of the panels.  The subband spacing is $\delta\varepsilon\approx21$~meV and the angle $\theta$ is taken from the $\hat{\vec z}$ direction. As expected from our analytical calculations, two shapes of non-uniform part of the chemical potential, $\delta\mu=\frac{3 \delta\varepsilon z}{L_z}$ and $\delta\mu=\frac{3 }{2} \delta\varepsilon \cos \theta$,  result in similar values of the splitting for the lowest angular momentum subbands. In contrast to that, for $\delta\mu=\frac{3 }{2 } \delta\varepsilon \cos 3\theta$, the splitting is weaker for the lowest $\ell=\pm 1/2$ subbands than in the first two cases, as also can be expected from our analytical treatment.}\label{sfig:Splittings}
\end{figure}

We use this model to confirm the validity of our perturbative solution for the size of the band splitting between two degenerate subbands as outlined in Eq.~\eqref{seq:bands}, see  Fig.~\ref{sfig:splittingstr}. We calculate the energy spectrum for three different strengths of the non-uniform part of the chemical potential: $\delta \mu (z) =n\delta\varepsilon z/L_z $ with $n=1,3,5$. We find that Eq.~\eqref{seq:bands} stays valid even when the applied bias is larger than the subband spacing $\delta\varepsilon$ and  deviates from the perturbative result only when the splitting energy of a given subband gets too large $E_{\rm so}\gtrsim\delta\varepsilon/2$.

To confirm that the splitting between degenerate subbands described by Eq.~\eqref{seq:bands} is not strongly dependent on the exact  form of the inhomogeneity of chemical potential and  is only sensitive to the relative sizes of the harmonics $\mu_n$, we consider three different chemical potential profiles $\delta \mu (z)$: $3 \delta\varepsilon z/L_z $ (as in the main text), $\frac{3}{2} \delta\varepsilon \cos \theta$, and $\frac{3}{2}  \delta\varepsilon \cos 3\theta$, with $\theta={\rm arctan}(x/z)$. The results are shown in Fig.~\ref{sfig:Splittings}, as expected, for the first two chemical potential profiles we find virtually no difference between the magnitude of splitting for the lowest subbands. For the third profile, the splitting is smaller for $\ell=\pm 1/2$ subbands than in the first two cases but approximately the same for  $\ell=\pm 3/2$, as expected from our analytic treatment. In all cases,  the splitting is well described by Eq.~\eqref{seq:bands}.

\subsection{Superconducting BHZ model} \label{app:tb}

Next, we add an on-site pairing amplitude $\Delta_{n,m}$, induced by proximity to an $s$-wave superconductor, to the Hamiltonian in momentum space defined by Eq.~(\ref{seq:BHZ}) and obtain 
\begin{align}
H_{\rm sc}(k)=&\sum_{n=1,m=1}^{L_x/a_1,L_z/a_3} \bar{\bf c}^\dagger_{n,m,k}\cdot\left[\{M(k)\tau_z+\frac{A_2}{a_2} \sin(k a_2) \tau_x \sigma_x  -\mu_{n,m} \}\eta_z+\Delta_{n,m}\eta_x\right]\bar{\bf c}_{n,m,k} \nonumber\\
+&\sum_{n=1,m=1}^{L_x/a_1-1,L_z/a_3} \left\{\bar{\bf c}^\dagger_{n+1,m,k} \cdot \left\{\frac{B_1}{a_1^2}\tau_z + \frac{i A_1}{2 a_1} \tau_x   \sigma_z \right\}\eta_z e^{i\eta_z \phi^x_{m} }  \bar{\bf c}_{n,m,k} +\text{H.c.} \right\} \nonumber \\
+&\sum_{n=1,m=1}^{L_x/a_1,L_z/a_3-1} \left\{\bar{\bf c}^\dagger_{n,m+1,k} \cdot \left\{\frac{B_3}{a_3^2}\tau_z + \frac{i A_3}{2 a_3} \tau_y \right\}\eta_z e^{i \eta_z \phi^z_{n} } \bar{\bf c}_{n,m,k}+\text{H.c.}\right\} , \label{seq:BHZSCmom}
\end{align}
where $\Delta_{n,m}=\Delta_0 e^{(m a_3-L_z)/\xi_{TI}}$ is the proximity induced pairing amplitude. We have introduced the extra set of Pauli matrices $\eta_{x,y,z}$ that act in particle-hole space of the Nambu operator and we have defined
\begin{equation}
{\bf \bar c}^\dagger_{n,m,k}=(c^\dagger_{+,\uparrow},c^\dagger_{-,\uparrow},c^\dagger_{+,\downarrow},c^\dagger_{-,\downarrow},c_{+,\uparrow},c_{-,\uparrow},-c_{+,\downarrow},-c_{-,\downarrow})_{n,m,k}
\end{equation}
in this space. This  superconducting BHZ model defined in momentum space is used for Fig.~3 of the main text.

To realise MBSs at the end of the wire as in Fig.~4 of the main text, we perform  the full three-dimensional BHZ tight binding model simulations:
\begin{align}
H_{\rm obc}=& \sum_{l=1,n=1,m=1}^{L_y/a_2,L_x/a_1,L_z/a_3} \bar{\bf c}^\dagger_{l,n,m}\cdot\left[\{M\tau_z -\mu_{n,m} \}\eta_z+\Delta_{n,m}\eta_x\right]{\bf c}_{l,n,m}\nonumber \\
&+\sum_{l=1,n=1,m=1}^{L_y/a_2-1,L_x/a_1,L_z/a_3} \left\{\bar{\bf c}^\dagger_{l+1,n,m} \cdot \left\{\frac{B_2}{a_2^2}\tau_z + \frac{i A_2}{2 a_2} \tau_x   \sigma_x \right\}\eta_z {\bf c}_{l,n,m} +\text{H.c.} \right\} \nonumber \\
&+\sum_{l=1,n=1,m=1}^{L_y/a_2,L_x/a_1-1,L_z/a_3} \left\{\bar{\bf c}^\dagger_{l,n+1,m} \cdot \left\{\frac{B_1}{a_1^2}\tau_z + \frac{i A_1}{2 a_1} \tau_x   \sigma_z \right\}\eta_z e^{i\eta_z \phi^x_{m} }  {\bf c}_{l,n,m} +\text{H.c.} \right\} \nonumber \\
&+\sum_{l=1,n=1,m=1}^{L_y/a_2,L_x/a_1,L_z/a_3-1} \left\{\bar{\bf c}^\dagger_{l,n,m+1} \cdot \left\{\frac{B_3}{a_3^2}\tau_z + \frac{i A_3}{2 a_3} \tau_y \right\}\eta_ze^{i \eta_z \phi^z_{n} } {\bf c}_{l,n,m}+\text{H.c.}\right\},  \label{3DBHZSC} 
\end{align}
with $M=M_0+2 \left(\frac{B_1}{a_1^2}+\frac{B_2}{a_2^2}+\frac{B_3}{a_3^2}\right)$ and the operators defined in the Nambu space as
\begin{equation}
{\bf \bar c}^\dagger_{l,n,m}=(c^\dagger_{+,\uparrow},c^\dagger_{-,\uparrow},c^\dagger_{+,\downarrow},c^\dagger_{-,\downarrow},c_{+,\uparrow},c_{-,\uparrow},-c_{+,\downarrow},-c_{-,\downarrow})_{l,n,m}.
\end{equation} In order to easily numerically simulate real system sizes of Bi$_{2}$Se$_{3}$, such as our $20$~nm~$\times~20$~nm~$\times~1000$~nm wire, whilst also minimising lattice effects, we adjust the Fermi velocity parallel to the wire which reduces the number of sites required along the wire. This is done via $\alpha$ in the parameter $A_2=\alpha 3.33$ eV\AA \ of our BHZ model. For the phase diagram in Fig. 4a and the disordered wires in Fig.~\ref{sfig:disorder} we choose $\alpha=1/5$ such that wires are $12\times24\times100$ lattice sites and for Fig. 4b we choose $\alpha=10/3$ such that wires are $12\times24\times150$ lattice sites. This change in $A_2$ (or equivalently the Fermi velocity $A_2/
\hbar$)  from the value with $\alpha=1$ has no influence on our results.

\begin{figure}[t]
\centering
\includegraphics[width=1\columnwidth]{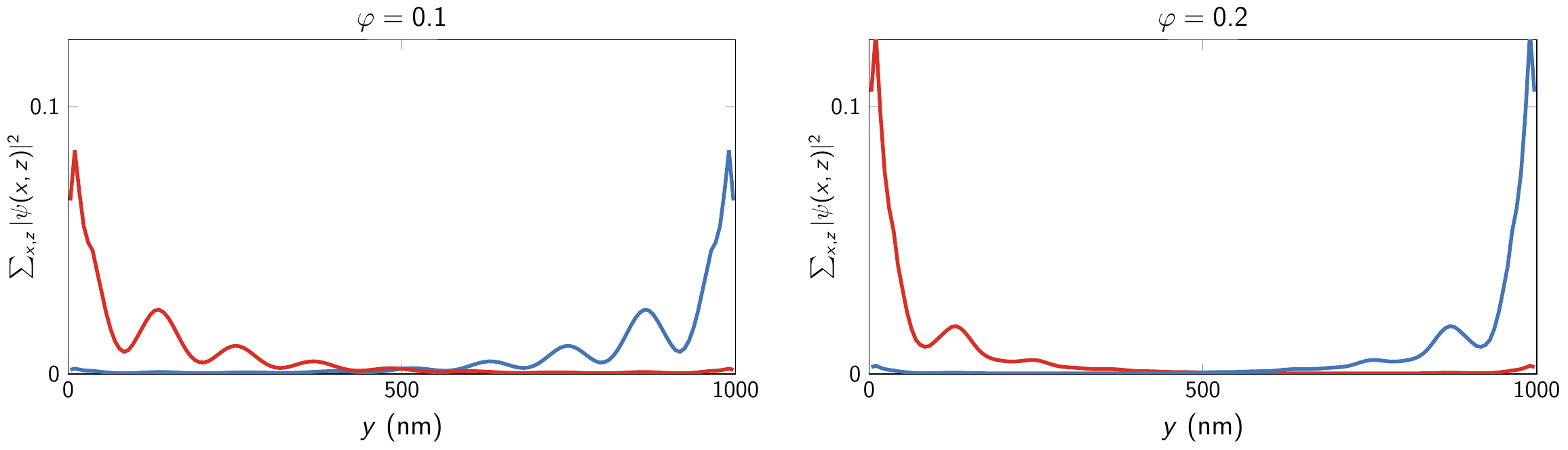}
\caption{{\bf Probability densities of MBSs along the wire:} The probability density $\sum_{x,z} |\psi(x,z)|^2$ of the lowest energy states (MBSs), summed over the cross-section of the TI wire, is calculated using the full three-dimensional BHZ model that includes also a superconducting pairing term [see Eq.~\eqref{3DBHZSC}]. 
The MBSs are localised at the ends of the wire. The localization length shrinks as magnetic flux is being increased.  For this system, as in the main text in Fig. 4, the critical flux value  $\varphi_c\approx0.065$. The wave functions of the two zero-energy MBSs are anti-symmetrised  such that they are maximally localised on the left (red) or right (blue). The other parameters are the same as in Fig. 4 of the main text.}\label{sfig:wvfs}
\end{figure}

When the topological criterion of the main text is fulfilled, the lowest energy states, MBSs, are close to zero in energy and  are well localised. The probability densities of the MBS wave functions along the wire, summed over the square cross-section of the TI wire, 
at two different fluxes are shown in Fig.~\ref{sfig:wvfs}.

\section{Disorder}
\begin{figure}[b]
\centering
\includegraphics[width=0.65\columnwidth]{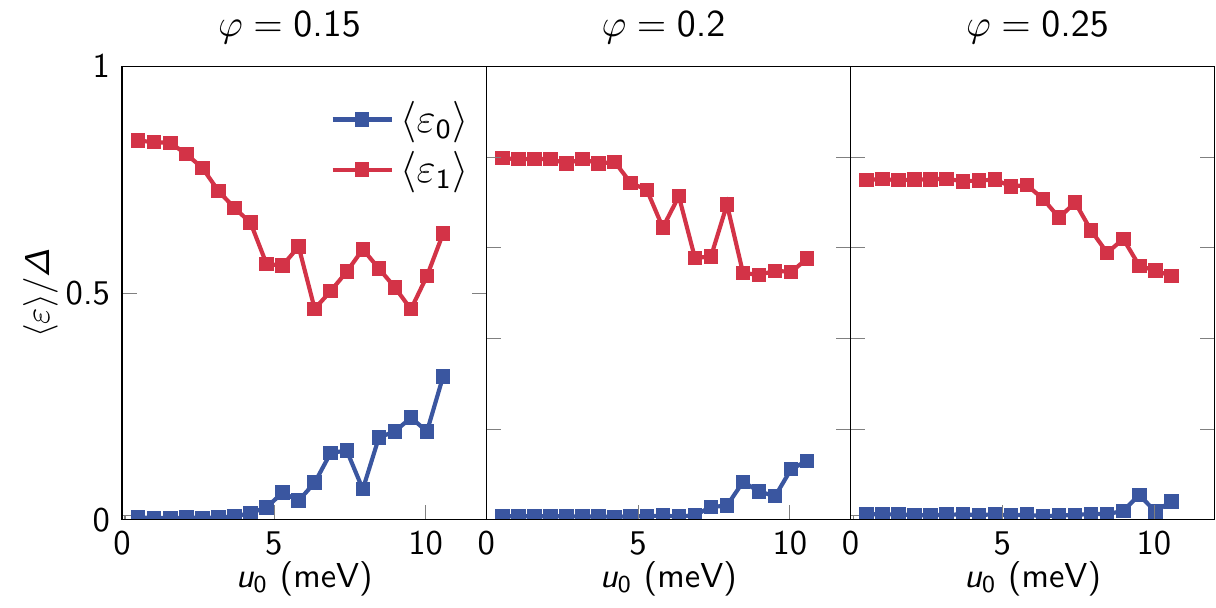}
\caption{{\bf The lowest and first excited energies for disordered wires:} We take the average over all lowest state energies, $\varepsilon_0$ and the average over all first excited state energies, $\varepsilon_1$, in an ensemble of 30 wires for various strengths of disorder $u_0$ and flux $\varphi$. The full three-dimensional BHZ model includes a superconducting pairing term [see Eq.\eqref{3DBHZSC}] with additional random on-site potential $\Delta\mu(\vec r)$. Even very strong disorder does not substantially move the MBS energy away from zero. For instance, for the largest value of flux in this plot, $\varphi=0.25$, disorder strengths up to $u_0\sim 8$ meV do not strongly affect the zero energy mode and it remains well separated from the SC bulk states. The other parameters are the same as in Fig. 4 of the main text.}\label{sfig:disorder}
\end{figure}

The appearance of MBSs that are pinned to zero energy for a large region of parameters indicates that MBSs in gated TI wires can be expected to be extremely stable against various types of disorder. This is important because present day bulk insulating TI wires are relatively dirty and transport in them is diffusive \cite{Munning2021}. In order to determine the stability of MBSs, we introduce on-site fluctuations in the chemical potential $\Delta \mu(\vec r)$ that are drawn randomly from a uniform distribution $\Delta \mu \in \left[-u_0/2,u_0/2\right]$. We take the average over all lowest state energies, $\varepsilon_0$, and the average over all energies of the first excited state, $\varepsilon_1$, in an ensemble of 30 wires with different disorder configurations at various strengths of $u_0$, the results are shown in Fig.~\ref{sfig:disorder} for various fluxes $\varphi$.  We find that the lowest energy state, corresponding to the MBSs, remains strongly pinned to zero energy and well separated from the SC bulk for very large disorder strengths. For instance, the ensemble-averaged lowest energy remains at zero up to $u_0\sim 8$ meV for $\varphi=0.25$ ($B\approx2.5$ T). This stability against disorder is a direct consequence of the exceptionally large phase space for MBSs in the topological phase diagram of our set-up.

In the context of disorder, thin wires are also more desirable as charged impurities in the bulk of thick TI samples can lead to long range $\sim 50$ nm fluctuations of the surface chemical potential up to several meV~\cite{skinner2013,skinner2013b,Borgwardt2016,Knispel2017} (so called surface puddles). The increased screening by the surface states in thin wires and the superconductor itself will likely significantly lower the magnitude and length of  surface puddles \cite{bomerich2017} and so, given the large area of topological phase in Fig.~4 of the main text, it is also likely that bulk charges will not affect the MBSs in our set-up.

\bibliography{TI-nws-si}